\documentstyle[prl,aps,psfig,epsfig,multicol]{revtex}
\begin{document}
\def\be{\begin{equation}}
\def\ee{\end{equation}}
\def\bc{\begin{center}}
\def\ec{\end{center}}
\def\bea{\begin{eqnarray}}
\def\eea{\end{eqnarray}}
\newcommand{\bleq}{\ifpreprintsty
                   \else
                   \end{multicols}\vspace*{-3.5ex}\widetext{\tiny
                   \noindent\begin{tabular}[t]{c|}
                   \parbox{0.493\hsize}{~} \\ \hline \end{tabular}}
                   \fi}
\newcommand{\eleq}{\ifpreprintsty
                   \else
                   {\tiny\hspace*{\fill}\begin{tabular}[t]{|c}\hline
                    \parbox{0.49\hsize}{~} \\
                    \end{tabular}}\vspace*{-2.5ex}\begin{multicols}{2}
                    \narrowtext \fi}

%\draft \twocolumn[\hsize\textwidth\columnwidth\hsize\csname
%@twocolumnfalse\endcsname
\title{Quantum statistics in complex networks}
\author{Ginestra Bianconi}
\address{Department of Physics, University of Notre Dame, Notre Dame,IN
 46556,USA }
\maketitle

\begin{abstract}
In this work we discuss the symmetric construction of bosonic and
fermionic networks and we present a case of a network showing a
mixed quantum statistics. This model takes into account the
different nature of nodes, described by a random parameter that we
call energy, and includes rewiring of the links. The system
described by the mixed statistics is an inhomogeneous system
formed by two class of nodes. In fact  there is a threshold energy
$\epsilon_s$ such that nodes with lower energy
$(\epsilon<\epsilon_s)$ increase their connectivity while nodes
with higher energy $(\epsilon>\epsilon_s)$ decrease their
connectivity in time.
\end{abstract}

\begin{multicols}{2}
\narrowtext
%\vskip2pc]

%\tableofcontents

\section{Introduction}
Recently, pushed by the need to fit the available experimental
data on a large variety of networks, statistical physics is
addressing its attention to complex
networks\cite{RMP,Strogatz,Doro_rev} and in particular to
scale-free networks characterized by power-law connectivity
distribution. The topological properties of these networks are
related to their dynamic evolution and play a key role in
collective phenomena of complex
systems\cite{Havlin,Vespignani,Stauffer}. Consequently there is an
urgent need of a general formalism  able to make a distinction
between networks. Different approaches have already been proposed
for equilibrium graphs\cite{Burda,Doro_stat}.

In this paper we will restrict our study to inhomogeneous growing
networks with different quality of nodes, described by quantum
statistics. In fact we have recently presented a growing
scale-free network with different qualities of the nodes and a
thermal noise that is described by Bose statistics\cite{bose}. On
the other hand we have found that a growing Cayley-tree with
different qualities of the nodes and a thermal noise is described
by Fermi statistics\cite{cayley-tree,IntJ}. In order to be
synthetic in the following we will refer to these two networks as
the bosonic and the fermionic networks respectively. Given the
fact that the solution of the dichotomy between Bose and Fermi
statistics is an attractive topic discussed in many different
contexts, from supersymmetry\cite{susy} to quantum algebras
\cite{q_algebras}, in the first part of the paper we compare the
growth dynamics of the two networks. We find that the bosonic and
fermionic network are obtained by continuous subsequent addition
of an elementary fan-shaped unit  attached in  two opposite
directions.
 While
the appearance of a classical system described by quantum
statistics is not completely new\cite{evans,lebowitz} this is the
first example of the occurrence of two symmetrically constructed
models following Bose and Fermi statistics respectively. Always
having in mind the general problem of the Bose-Fermi dichotomy, in
the second part of this work we provide a more realistic example
of network in which the two growth processes coexist. This is
obtained by rewiring a bosonic network. This complex inhomogeneous
system has
 two classes of nodes with increasing and decreasing
connectivity and is fully described by a mixed statistic depending
on two chemical potentials ($\mu_B$ and $\mu_F$).

\section{Symmetric construction of bosonic and fermionic networks}

The bosonic network \cite{bose} is a scale-free network in which
each node has an intrinsic quality $\epsilon$
  from a time
independent distribution $p(\epsilon)$. At each timestep a new
node is added to the network attaching $m$ links preferentially to
more connected low energy nodes. The probability $\Pi_i$ that a
new link is attached to a node of energy $\epsilon_i$ and
connectivity $k_i$ is given by
\be
\Pi_i^B\propto e^{-\beta \epsilon_i}k_i. \label{uno}\ee The
fermionic network \cite{cayley-tree} is a growing Cayley tree of
coordination number $m+1$ in which nodes have an intrinsic quality
$\epsilon$ from a time independent distribution $p(\epsilon)$.
Nodes are distinguished between nodes at the interface (with
connectivity one) and nodes in the bulk (with connectivity $m+1$).
At each time step a node at the interface can grow giving rise to
$m$ new nodes. The probability that a node $i$ grows is given by
the probability $\rho_i$ that the node is at the interface   (its
survivability), times $e^{\beta\epsilon_i}$
\be
\Pi_i^F\propto e^{\beta \epsilon_i} \rho_i. \label{due}\ee
 The
dynamic of the two networks is parameterized by $\beta$ that is a
characteristic of the network growth and plays the role of the
inverse temperature, i.e. $\beta=1/T$. For $T=0$ the dynamics
became extremal and $\Pi_i^B$, $\Pi_i^F$ are different form zero
only for the  lowest and the highest energy node of the network
respectively. As the temperature increases, the dynamic involves
also the other nodes and in the $T\rightarrow \infty$ limit
$\Pi_i^B$ and $\Pi_i^F$ do not depend anymore on the energy of the
nodes.

 A generic bosonic network following $(\ref{uno})$ and a
generic fermionic network following $(\ref{due})$ can be
constructed by attaching a fixed elementary unit
  to a   number of nodes growing linearly with the size of the network $N$.

The fixed elementary unit playing the role of the
 'unitary cell'
in crystal lattices, is a fan-shaped element constituted by a
vertex node connected to $m$ other nodes. But the way in which
this unit is attached is symmetric in the two networks. In the
bosonic network the vertex of the fan is a new node attached  by
$m$ links to $m$ of the $N$ existing nodes of the network. On the
contrary, in the fermionic network the  elementary unit is
reversed and the vertex is one of the $(1-1/m)N$ nodes at the
interface while the  $m$ nodes attached to it are new nodes of the
network. Consequently both networks are constructed by the
addition of the same elementary unit attached in the two opposite
directions.

The mean-field equation for the bosonic and fermionic network
describe respectively the evolution of the connectivity $k_i$ and
the survivability $\rho_i$ of the nodes. In the  bosonic network,
since  every new link is attached to node $i$ with probability
$(\ref{uno})$, and $m$ new link are attached at each timestep, the
mean field equation for the connectivity $k_i$ is given by
\be
\frac{\partial k_i}{\partial t}=m\frac{e^{-\beta
\epsilon_i}k_i}{\sum_j e^{-\beta \epsilon_j}k_j } \label{ki}\ee
where $\sum_j e^{-\beta \epsilon_j}k_j $ is the normalization sum
of the probability $\Pi_i^B$, Eq. $(\ref{uno})$.
 Symmetrically, in a fermionic network every node grows with
probability $\Pi_i^F$ given by $(\ref{due})$. Consequently the
probability $\rho_i(t)$ that a node $i$ is at the interface
decreases in time following the mean field equation
\be
\frac{\partial \rho_i}{\partial t}=-\frac{e^{\beta
\epsilon_i}\rho_i}{\sum_j e^{\beta \epsilon_j}\rho_j }
\label{rhoi}\ee where the denominator sum is needed in order to
normalize the probability $\Pi_i^F$. In both networks the
resulting structure optimizes the system by minimizing the 'free
energy' of each node of the network $\epsilon_i-T \log(k_i)$
(bosonic network) or $\epsilon_i-T |\log \rho_i|$ (fermionic
network). In the two networks this optimization is achieved in
different ways. In the bosonic network the low-energy nodes are
more likely to be awarded a new link while in the fermionic
network high-energy nodes are more likely to be removed from the
interface. While geometrically the two networks are related by the
reversal of the elementary unit, the mean field equations
(\ref{ki}) and $(\ref{rhoi}$), in the case $m=1$, are symmetric
under  time reversal ($t\rightarrow-t$) and the change of sign of
the energies ($\epsilon_i\rightarrow -\epsilon_i$).
Self-consistent calculations $\cite{bose,cayley-tree}$ show that
the connectivity $k(t|\epsilon,t')$ (the survivability
$\rho(t|\epsilon,t'$)) of a node of energy $\epsilon$ added to the
network at time $t'$, follows a power-law in time with an exponent
dependent on its energy, \bea
k(t,|\epsilon,t')=m\left(\frac{t}{t'}\right)^{f_B(\epsilon)}
&{\mbox{with  }}& {f_B(\epsilon)=e^{-\beta(\epsilon-\mu_B)}},
\nonumber
\\ \rho(t,|\epsilon,t')=\ \ \left(\frac{t'}{t}\right)^{f_F(\epsilon)}
& {\mbox{with  }}    & {f_F(\epsilon)=e^{\beta(\epsilon-\mu_F)}}.
\label{power}\eea The time reversal of the two mean-field solution
implies here that the connectivity of the nodes always increases
in time in the bosonic networks while the survivability of the
nodes always decreases in time in the fermionic network. The
dynamic described by $(\ref{power}$) depends on the two constant
 $\mu_B$ and $\mu_F$ given, respectively, by
the solutions of the two equations
 \bea 1&=\int{d\epsilon p(\epsilon)
\frac{1}{e^{\beta(\epsilon-\mu_B)}-1}}&=\int d\epsilon p(\epsilon)
n_B(\epsilon) ,
 \nonumber \\
1-\frac{1}{m}&=\int{d\epsilon p(\epsilon)
\frac{1}{e^{\beta(\epsilon-\mu_F)}+1}}&=\int d\epsilon p(\epsilon)
n_F(\epsilon), \label{mu}\eea where $n_B/n_F(\epsilon)$ indicates
the bosonic and fermionic occupation numbers respectively. Thus
the evolution of each node of the network is completely determined
by a number, $\mu_B$ or $\mu_F$, defined as the chemical
potentials of a bosonic or fermionic system with specific volumes
$v_B=1$ and $v_F=1+1/(m-1)$ respectively.

The quantum occupation numbers $n_B(\epsilon)$ and $n_F(\epsilon)$
appears spontaneously in the solution of the mean-field equations
$(\ref{ki})$ and ($\ref{rhoi}$) and assume a clear meaning when we
look at the static picture of the networks.
 In fact, in the bosonic network the
 total number of links attached to nodes with energy $\epsilon$,
$N_B(\epsilon)$ is given by \be N_B(\epsilon)=m t
p(\epsilon)[1+n_B(\epsilon)]. \label{nbs}\ee In the l.h.s. of
Eq.~$(\ref{nbs})$ the first and second terms represent the number
of outgoing and incoming links connected to  nodes of energy
$\epsilon$. Similarly,  in the fermionic network, the total number
of nodes with energy $\epsilon$ found below the interface,
$N_F(\epsilon)$ is given by the difference between all the nodes
of the network and those that are at the interface, i.e.
\be
N_F(\epsilon)=m t p(\epsilon)[1-n_F(\epsilon)]. \label{nfs}\ee

 Nevertheless $n_B(\epsilon)$ and $n_F(\epsilon)$
 acquire also a very specific role in the single time evolution
of the network.  In fact, at time $t$, the probability
$\pi_B^{(t)}(\epsilon)$ of attaching a new link to a generic node
of energy $\epsilon$ (bosonic network) and the probability
$\pi_F^{(t)}(\epsilon)$ that a generic node with energy $\epsilon$
will grow in the fermionic network, are given  by \bea
\pi_B^{(t)}(\epsilon)&=\int_1^t dt' \delta(\epsilon-\epsilon_{t'})
\frac{\partial k(t|\epsilon_{t'},t')}{\partial t}& \rightarrow
p(\epsilon) n_B(\epsilon) \nonumber \\
\pi_F^{(t)}(\epsilon)&=\int_1^t dt' \delta(\epsilon-\epsilon_{t'})
\frac{\partial \rho(t,|\epsilon_{t'},t')}{\partial t}& \rightarrow
p(\epsilon) [1-n_F(\epsilon)]. \eea These results explain the
interconnection between the dynamic of the networks and their
self-similar aspect. In fact,  for the bosonic network we have
that  the probability for a new node to be linked to a node with
energy $\epsilon$ converges in time to the same limit as the
density of existing links pointing to nodes of energy $\epsilon$.
Similarly, for the fermionic network we have that the probability
that a node with energy $\epsilon$ is chosen to grow converge to
the same limit than the density of nodes in the bulk.

The occurrence of the two quantum statistics
 in the description of such networks is due  to the fact that the
 networks are growing by the continuous addiction of the unitary
 cell but they try also to minimize the energy of the system (by the choice of the
 node to which attach a new link in the bosonic network
 or by the choice of the growing
 node in the fermionic network).
The stochastic model behind the construction of the two networks
always involves the choice of a node in between a growing number
of nodes, but while in the Cayley tree a chosen node is removed
from the interface and cannot be chosen any more, in a scale-free
networks there is no limit to the number of links a node can
acquire. Consequently the Cayley tree is described by a Fermi
distribution while the scale-free network is described by a Bose
distribution.

The framework of quantum statistics clarify the relation between
self-organized critical processes and scale-free models. In fact,
the  fermionic network evolution in
 the $T \rightarrow 0$ limit reduces to the Invasion
Percolation dynamics on a Cayley tree\cite{Thorpe,IPCT,Gabrielli},
a well known self-organized process\cite{BTW} while the bosonic
network in the $T\rightarrow\infty$ limit reduces to the BA
model\cite{BA} for growing scale-free networks.

\section{Mixed statistics in scale-free network with rewiring}

 Our purpose here is to expand
on the previous results and to discuss systems which are governed
by additional processes on top of the simple growth discussed
before. For example, in real networks, in addition of the
appearance of new nodes, one can observe new links as well, or
rewiring of existing links. In fact rewiring of the link in a
scale-free network has been used to model increasing disorder in
more realistic networks \cite{rewiring,Capocci}. We show that the
presence of such additional processes can create a coexistence of
 Fermi and Bose statistics within the same system.
This implies that most real systems, for which such additional
processes are present, exist in a mixed state, whose statistics
can be described only by simultaneously involving both Bose and
Fermi statistics. It is not our purpose to model any particular
system at this point. Thus next we discuss a simple system that
displays this mixed behavior.

A simple example of mixed statistics is given by introducing
rewiring into a bosonic  network. This network
 is constructed iteratively in the following way: at each timestep a
new node and $m$ links are added to the network. The new node has
an energy $\epsilon$ chosen from a distribution $p(\epsilon)$ and
the $m$ links connect the new node preferentially to well
connected, low energy nodes of the system. As in the bosonic
network without rewiring we assume that a new links is attached
with probability
\be
\Pi^+_{i}\propto e^{-\beta \epsilon_i}k(t|\epsilon_i,t_i) \ee to
node $i$ arrived in the network at time $t_i$, with energy
$\epsilon_i$ and connectivity $k(t|\epsilon_i,t_i)$ at time $t$.
Furthermore we assume also that at each timestep $m'$ edges detach
from existing nodes and are rewired to the new node. Consequently
every new node will have $m+m'$ links attached to it. We assume
that edges connected to high energy nodes are more unstable, so
that the probability that an edge connected to a node of energy
$\epsilon_i$ detaches from it is proportional to $e^{\beta
\epsilon_i}$. Consequently, the probability that a node $i$ will
loose a link because of the rewiring is given by
\be
\Pi^-_{i}\propto e^{\beta \epsilon_i}k(t|\epsilon_i,t_i) \ee where
$t_i$ is the time node $i$ is added in the network , $\epsilon_i$
is its energy and $k(t|\epsilon_i,t_i)$ its connectivity at time
$t$.
The continuous equation describing the time evolution of the
connectivities of the nodes is given by
\be
\frac{\partial k(t|\epsilon_i,t_i)}{\partial t}=m\frac{e^{-\beta
\epsilon_i}k(t|\epsilon_i,t_i)}{\sum_j e^{-\beta
\epsilon_j}k(t|\epsilon_j,t_j)}-m'\frac{e^{\beta
\epsilon_i}k(t|\epsilon_i,t_i)}{\sum_j e^{\beta
\epsilon_j}k(t|\epsilon_j,t_j)} \label{dyn.eq} \ee with the
initial condition
\be
k_0=k(t|\epsilon_i,t)=m+m'. \label{in.eq}\ee To solve
$(\ref{dyn.eq})$ we assume that in the thermodynamic limit the
normalization sums $Z_B$ and $Z_F$, given by \bea Z_B&=&\sum_j
e^{-\beta\epsilon_j}k(t|\epsilon_j,t_j) \nonumber
\\Z_F&=&\sum_j
e^{\beta\epsilon_j}k(t|\epsilon_j,t_j) \eea selfaverage and
converge to their mean value $<Z_B>_{\epsilon}$ and grow linearly
in time,with the asymptotic behavior given by the constants
$\mu_B$ and $\mu_F$, $<Z_F>_{\epsilon}$, \bea
Z_B\rightarrow<Z_B>_{\epsilon}&\rightarrow&mt e^{-\beta
\mu_B}\nonumber \\ Z_F\rightarrow<Z_F>_{\epsilon}&\rightarrow&m't
e^{\beta \mu_F}. \label{limit.eq} \eea  Using $(\ref{limit.eq})$,
the dynamic equation $(\ref{dyn.eq})$ reduces to
\be
\frac{\partial k(t|\epsilon_i,t_i)}{\partial t}=(e^{-\beta
(\epsilon_i-\mu_B)}- e^{\beta
(\epsilon_i-\mu_F)})\frac{k(t|\epsilon_i,t_i)}{t}. \ee
Consequently we have found that the time evolution of the
connectivity $k(t|\epsilon_i,t_i)$ follows a power-law
\be
k(t|\epsilon,t')=k_0\left(\frac{t}{t'}\right)^{f_{mix}(\epsilon)}
\label{kt.eq} \ee with
\be
f_{mix}(\epsilon)=e^{-\beta(\epsilon-\mu_B)}-e^{\beta(\epsilon-\mu_B)}.
\label{f.eq}\ee

 The characteristic difference of this network from the bosonic
scale-free network is that the connectivity of the nodes, due to
the rewiring process, can either increase or decrease in time. In
fact $f_{mix}(\epsilon)$ (defined in Eq. ($\ref{f.eq}$)) change
sign at a threshold energy value
\be
\epsilon_s=\frac{\mu_B+\mu_F}{2}. \ee Consequently the nodes with
energy $\epsilon<\epsilon_s$ increase their connectivity in time
while nodes with energy higher then the threshold $\epsilon_s$,
i.e. $\epsilon>\epsilon_s$, have a decreasing connectivity.

After substituting $k(t|\epsilon_i,t_i)$ form Eq. $(\ref{kt.eq})$
with $f_{mix}(\epsilon)$ given by $(\ref{f.eq})$ into
$(\ref{dyn.eq})$ and the sum over $j$ with an integral , we get
the self-consistent equations
 for the chemical potential $\mu_B$ and $\mu_F$
  \bea \frac{m}{m+m'}&=& \int d\epsilon p(\epsilon)
\frac{e^{-\beta(\epsilon-\mu_B)}}{1-e^{-\beta(\epsilon-
\mu_B)}+e^{\beta(\epsilon-\mu_F)}}\nonumber
\\ \frac{m'}{m+m'}&=&\int d\epsilon p(\epsilon) \frac{e^{\beta(\epsilon-\mu_F)}}{1-e^{-\beta(\epsilon-\mu_B)}+e^{\beta(\epsilon-\mu_F)}}
. \label{Rew.fmix}\eea

\begin{figure}
\centerline{\epsfxsize=3.5in \epsfbox{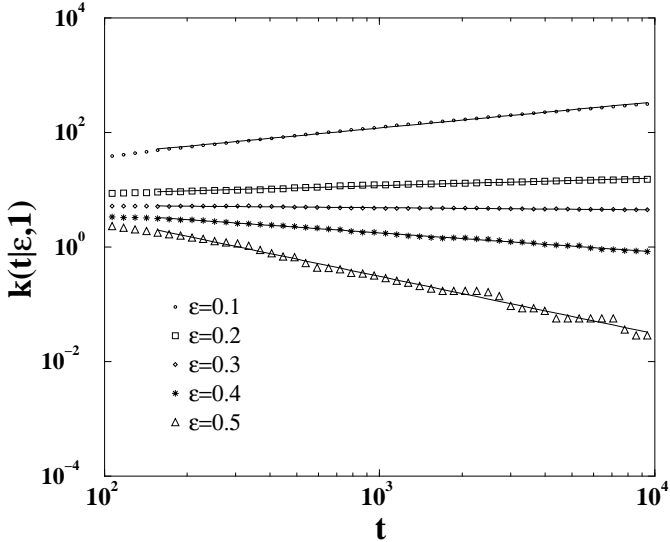}}
\caption{Dynamical evolution of the connectivity of nodes with
different energies. The connectivity of the nodes always follows a
power-law, increasing or decreasing in time  depending on the
energy $\epsilon$ and the threshold value $\epsilon_s$.}
\label{fig3}
\end{figure}

On the same time, the distribution of edges attached to  nodes
with energy $\epsilon$ converges to the mixed statistics
\be
n_{mix}(\epsilon)p(\epsilon)=(m+m')p(\epsilon)\frac{1}{1-e^{-\beta
(\epsilon-\mu_B)}+e^{\beta (\epsilon-\mu_F)}} \ee while the number
$n_{+}(\epsilon)p(\epsilon)$ of the edges stochastically attached
to the nodes of energy $\epsilon$ or the number
$n_{-}(\epsilon)p(\epsilon)$ of the nodes detached from nodes with
energy $\epsilon$ are given by \bea n_{+}(\epsilon)p(\epsilon)&=&
m p(\epsilon)\frac{e^{-\beta (\epsilon-\mu_B)}}{1-e^{-\beta
(\epsilon-\mu_B)}+e^{\beta (\epsilon-\mu_F)}}, \nonumber \\
n_{-}(\epsilon)p(\epsilon)&=& m' p(\epsilon)\frac{e^{\beta
(\epsilon-\mu_F)}}{1-e^{-\beta (\epsilon-\mu_B)}+e^{\beta
(\epsilon-\mu_F)}} \eea respectively. The distribution
$n_{mix}(\epsilon)$ appears as a natural candidate of a mixed
statistics going from the $\mu_F\rightarrow \infty$ limit where
$n_{mix}(\epsilon)\propto 1+n_B(\epsilon)$ to the
$\mu_B\rightarrow \infty$ limit where $n_{mix}(\epsilon)\propto
n_F(\epsilon)$.

 We have simulated a network with $m=2$ and $m'=1$ and uniform energy
distribution $p(\epsilon)=1$ for $\epsilon \in (0,1)$, with
chemical potentials $\mu_B=0.03$, $\mu_F=0.51$ and
$\epsilon_s=0.27$. In Fig. $\ref{fig3}$ we show the connectivity
of the nodes of the network with energy values above and below the
threshold $\epsilon_s=0.27$. The figure shows that nodes with
energy $\epsilon<\epsilon_s$ increase their connectivity in time
while nodes with energy $\epsilon>\epsilon_s$ decrease their
connectivity in time. In Fig.  $\ref{fig4}$ we report the number
of links attached to nodes of energy $\epsilon$,
$n_{mix}(\epsilon)$ for a system size $N=10^4$ with the data
averaged over $100$ runs. In the same figure we report also the
number of nodes stochastically attached to (detached from)  nodes
of energy $\epsilon$, $n_+(\epsilon)$ ($n_-(\epsilon)$).

\begin{figure}
\centerline{\epsfxsize=3.5in \epsfbox{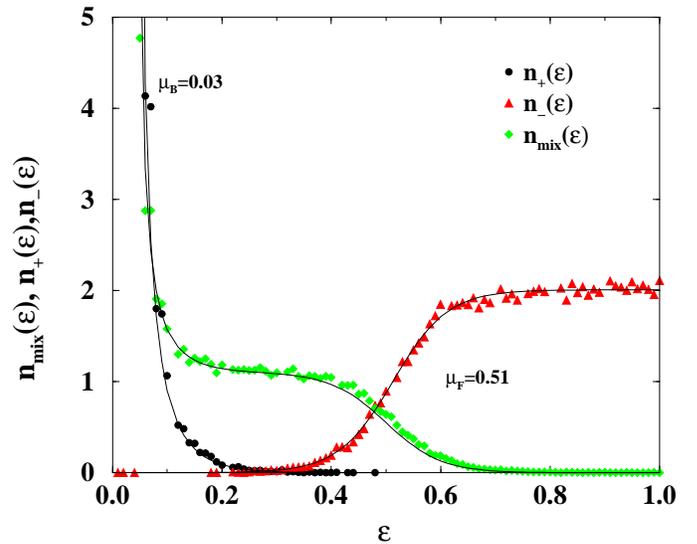}} \caption{ The
number $n_{mix}$ of edges attached to the nodes with energy
$\epsilon$, the number $n_{+}(\epsilon)$ of the edges
stochastically attached to  the nodes with energy $\epsilon$, the
number $n_{-}(\epsilon)$ of the nodes detached from nodes with
energy $\epsilon$ are plotted as a function of energy. The
simulations have been obtained with a uniform energy distribution
in the interval $[0,1]$. The data for $10^5$ timesteps are
averaged over 100 runs.} \label{fig4}
\end{figure}

The connectivity distribution $P(k)$ is given by the sum of the
probabilities $P(k|\epsilon)$ that a node with energy $\epsilon$
has connectivity $k$. Thus, if $k>m+m'$ we have to  sum over all
the nodes with energy lower then the threshold $\epsilon_s$, while
if $k<m+m'$ the summation will be over the nodes with energies
higher than the threshold, \bea P(k)&=&
\theta(k-k_0))\frac{t}{k_0}\int_{\epsilon<\epsilon_s} d\epsilon
p(\epsilon)\frac{1}{|f_{mix}(\epsilon)|}{
\left(\frac{k}{k_0}\right)}^{-\gamma(\epsilon)} + \nonumber \\
&+&\theta(k_0-k)  \frac{t}{k_0}\int_{\epsilon>\epsilon_s}
d\epsilon p(\epsilon) \frac{1}{|f_{mix}(\epsilon)|}{
\left(\frac{k}{k_0}\right)}^{-\gamma(\epsilon)}\eea with
\be
\gamma(\epsilon)=1+1/{f_{mix}(\epsilon)}
\ee
with
\bea
\gamma(\epsilon)>1 &\mbox{  for } \epsilon<\epsilon_s, \nonumber \\
\gamma(\epsilon)<1 &\mbox{  for } \epsilon>\epsilon_s.
\eea

In the {\it limit $\beta\rightarrow 0$} all the nodes of the network
evolve in the same way with

\be
f_{mix}(\epsilon)=\frac{m-m'}{2m}= \Delta. \ee Thus, if $\Delta>0$
every node increases its connectivity in time while if $\Delta<0$
all the nodes have decreasing connectivity. In the case $\Delta=0$
the mean field equation describes a system in which the
connectivities remain constant in time.

On the contrary in the
{\it limit  $ \beta \rightarrow \infty$}
the difference between nodes with different energy is strongly
enhanced.

We have to observe that as $\Delta $ goes from its highest value
$\Delta=1/2$ to negative values, the energy distribution goes from
a pure Bose distribution to a mixed distribution with an
increasing Fermi character, i.e. with a decreasing Fermi potential
$\mu_F$. But it is impossible to reach the pure Fermi statistics
in this way. In fact if we consider the limit $m=0$, the number of
links in the network is not increasing in time, and the new nodes
only acquire edges from the rewiring process. In this case the
connectivity of the nodes decreases exponentially as
\be
k(t|\epsilon,t_i)=k_0 \exp(-e^{-\beta (\epsilon-\mu_F)}(t-t_i))
\ee
with the chemical potential defined by
\be
N=\int d\epsilon p(\epsilon) m' e^{\beta (\epsilon-\mu_F)}. \ee We
observe that in this case the network doesn't grow anymore and the
number of edges attached to nodes with energy $\epsilon$ is simply
given by the Boltzmann occupation factor. In this case the
self-consistent equation and the mass conservation relation are
not anymore equivalent, the first one reducing, in the
thermodynamic limit to an identity. For this network the
probability $P(k)$ to find a node with connectivity $k$ is given
by
\be
P(k)=\frac{1}{k} \int d\epsilon p(\epsilon) k_0 e^{\beta(
\epsilon-\mu_F)}, \ee i.e. goes like $P(k) \sim k^{-1}$.

\section{Conclusions}
In conclusion we have shown the symmetry between the fermionic and
the bosonic networks emphasizing the role of quantum statistics.

These two particular evolving networks are related by the time
reversal evident in the continuum equations describing their
dynamics  and in the reversed unitary unit by which the two
network are built. This time reversal implies that the
connectivity increases in time while the survivability of each
node decreases in time as an energy-dependent power-law. The time
reversal of the single process generate two different structures
with properties and dynamics  only described
 by the functionals $\mu_B$ and $\mu_F$, at every temperature
$T=1/\beta$.  Having introduced these two limit simple cases and
having illustrated their symmetry we have shown that it is
possible to construct a new class of networks described by a mixed
statistics that can be applied to
 real systems where the two different growth  processes
coexist.

\section{Akwnoledgements}

 I am grateful to professor A.-L. Barab\'asi for useful
discussions and to professor G. Jona-Lasinio  and doctor R. S.
Johal for introducing me respectively to supersymmetry and quantum
algebras. This work was supported by NSF.

%%%%%%%%%%%%%%%%%%%%%%%%%%%%%% ACKNOWLEDGMENTS %%%%%%%%%%%%%%%%%%%%%%%%%%%%%%
%%%%%%%%%%%%%%%%%%%%%%%%%%%%%%%%%%%%%%%%%%%%%%%%%%%%%%%%%%%%%%%%%%%%%%%%%%%%%%

\end{multicols}
\end{document}